\documentstyle[epsfig,12pt]{article}  
\newcommand{\bce}{\begin{center}}
\newcommand{\ece}{\end{center}}
\newcommand{\beq}{\begin{equation}}
\newcommand{\eeq}{\end{equation}}
\newcommand{\bea}{\vspace{0.25cm}\begin{eqnarray}}
\newcommand{\eea}{\end{eqnarray}}

\newcommand{\ba}{\begin{array}}
\newcommand{\ea}{\end{array}}

\newcommand{\r}{\mbox{{\boldmath
$\rho$}}}
\newcommand{\ta}{\mbox{{\boldmath
$\tau$}}}

\newcommand{\qb}{\mbox{{\bf
q}}}

\newcommand{\rb}{\mbox{{\bf
r}}}

\def\lsim{\mathrel{\rlap{\lower4pt\hbox{\hskip1pt$\sim$}}
    \raise1pt\hbox{$<$}}}         %less than or approx. symbol
\def\gsim{\mathrel{\rlap{\lower4pt\hbox{\hskip1pt$\sim$}}
    \raise1pt\hbox{$>$}}}         %greater than or approx. symbol

    \def\beq{\begin{equation}}
    \def\endeq{\end{equation}}
    \def\bea{\begin{eqnarray}}
    \def\arr{\begin{eqnarray}}
    \def\eea{\end{eqnarray}}

\def\q2{$Q^{2}$}
\def\s2{2$S$}

\begin{document}
\thispagestyle{empty}
\vspace*{-2cm}
\begin{flushright}
{\bf\large
FZJ-IKP(Th)-1999-16\\}
%\,\,28 June 1999\,\,\,\,\,\,\,\,}
\end{flushright}
 
\bigskip

\begin{center}

  {\large\bf
TRANSVERSE SPECTRA OF 
RADIATION PROCESSES IN MEDIUM
\\
\vspace{1.5cm}
  }
\medskip
  {\large
  B.G. Zakharov
  \bigskip
  \\
  }
{\it  Institut  f\"ur Kernphysik,
        Forschungszentrum J\"ulich,\\
        D-52425 J\"ulich, Germany\medskip\\
 L.D. Landau Institute for Theoretical Physics,
        GSP-1, 117940,\\ Kosygina Str. 2, 117334 Moscow, Russia
%        \medskip\\
\vspace{2.7cm}\\}

  {\bf
  Abstract}
\end{center}
{
\baselineskip=9pt
We develop a formalism for evaluation of
the transverse momentum dependence of cross sections of the radiation processes
in medium.
The analysis is based on the light-cone path integral approach to
the induced radiation. The results are applicable in both QED and QCD.
}
\pagebreak
%-------------------------------------------------------------
\newpage
%-------------------------------------------------------------

It is well known that at high energies the multiple scattering can 
considerably modify cross sections of the radiation processes 
in medium \cite{LP,Migdal}.
Recently this effect (called the Landau-Pomeranchuk-Migdal (LPM) effect)
in QED and QCD has attracted much attention
\cite{BDPS,Knoll,BD,BGZ1,BGZ_SLAC,BDMPS,BGZ2,BDMS} 
(see also \cite{Klein} and references therein).
In \cite{BGZ1} we have developed a new rigorous light-cone 
path integral approach 
to the LPM effect.
There we have discussed
the $p_{T}$-integrated spectra.
For many problems it is highly desirable
to have also a formalism for the $p_{T}$-dependence
of the radiation rate. 
In the present paper we derive the corresponding formulas.
Similarly to \cite{BGZ1} our results are applicable 
in both QED and QCD.

For simplicity we describe the formalism  
for an induced $a\rightarrow bc$ transition in QED for scalar particles
with an interaction Lagrangian  
$L_{int}=\lambda [\hat\psi_{b}^{+}\hat\psi_{c}^{+}\hat\psi_{a}+
\hat\psi_{a}^{+}\hat\psi_{c}\hat\psi_{b}]$ 
(it is assumed that $m_{a}<m_{b}+m_{c}$, and the decay $a\rightarrow bc$
in vacuum is absent).
The $S$-matrix element for the $a\rightarrow bc$ transition 
in an external potential reads
\beq
\langle bc|\hat{S}|a\rangle=i\int\! dt
d\rb \lambda
\psi_{b}^{*}(t,\rb)\psi_{c}^{*}(t,\rb)\psi_{a}(t,\rb)\,,
\label{eq:1}
\eeq
where $\psi_{i}$ are the wavefunctions (incoming for $i=a$ 
and outgoing for $i=b,c$). We write $\psi_{i}$ as
\beq
\psi_{i}(t,\rb)=\frac{1}{\sqrt{2E_{i}}}\exp[-i(t-z)p_{i,z}]
\phi_{i}(t,\rb)\,.
\label{eq:2}
\eeq
We consider the case when the particle $a$ approaches the target from
infinity,
and normalize the flux to unity
(it corresponds to $|\phi_{i}|=1$) at $z=-\infty$ for $i=a$ 
and at $z=\infty$ for $i=b,c$.
The case when the particle $a$ is produced in a hard reaction in a medium
(or at finite distance from a medium) 
will be discussed later. 
At high energy, $E_{i}\gg m_{i}$,
the dependence of $\phi_{i}$ on the variable 
$\tau=(t+z)/2$ at $t-z=$const is governed by the two-dimensional 
Schr\"odinger equation
\beq
i\frac{\partial{\phi_{i}}}{\partial{\tau}}=
H_{i}\phi_{i}\,,
\label{eq:3}
\eeq
\beq
H_{i}=-\frac{1}{2\mu_{i}}
\left(\frac{\partial}{\partial\r}\right)^{2}
+e_{i}U+\frac{m_{i}^{2}}{2\mu_{i}}\,,
\label{eq:4}
\eeq
where $\mu_{i}=p_{i,z}$, 
$\r$ is the transverse coordinate,
$e_{i}$ is the electric charge, and $U$ 
is the potential of the target.

In the high energy limit from 
(\ref{eq:1}), (\ref{eq:2}) one can obtain 
for the inclusive cross section 
\beq
\frac{d^{5}\sigma}{dx d\qb_{b}d\qb_{c}}=\frac{2}{(2\pi)^{4}}
\mbox{Re}\!
\int\!d\r_{1}d\r_{2}
\!\!\int
\limits_{z_{1}<z_{2}}\!\!dz_{1}dz_{2}\,
g\langle F(z_{1},\r_{1})
F^{*}(z_{2},\r_{2})\rangle\,,
\label{eq:5}
\eeq
where 
$
\left.F(z,\r)=\phi_{b}^{*}(t,\rb)
\phi_{c}^{*}(t,\rb)\phi_{a}(t,\rb)\right|_{t=z}\,,
$
$\qb_{b,c}$ are the transverse momenta, 
$x=p_{b,z}/p_{a,z}$ (note that for the particle $c$  
$p_{c,z}=(1-x)p_{a,z}$),
$g=\lambda^{2}/[16\pi x(1-x)E_{a}^{2}]$,
$\langle ...\rangle $
means averaging over the states of the target.
Since the wavefunctions enter (\ref{eq:5}) only at $t=z$,
$\phi_{i}$ can be regarded 
as functions of $z$, and $\r$. 
In the Schr\"odinger equation (\ref{eq:3})
$z$ will play the role of time.
We represent the $z$-dependence of $\phi_{i}$ in terms of the 
Green's function, $K_{i}$, of the Hamiltonian (\ref{eq:4}).
Then, diagrammatically, (\ref{eq:5})
is described by the graph of Fig.\,1a. We depict 
$K_{i}$ ($K_{i}^{*}$) by $\rightarrow$ ($\leftarrow$).
The dotted line
shows the transverse density matrices at large longitudinal 
distances
in front of ($z=z_{i}$) and behind ($z=z_{f}$) 
the target.\footnote{Strictly speaking, in 
(\ref{eq:1}), (\ref{eq:5})
the adiabatically vanishing at $|z|\sim |z_{i,f}|$ coupling 
should be used. 
For simplicity we do not indicate
the coordinate dependence of the coupling.}
\begin{figure}[h]
\begin{center}
\psfig{file=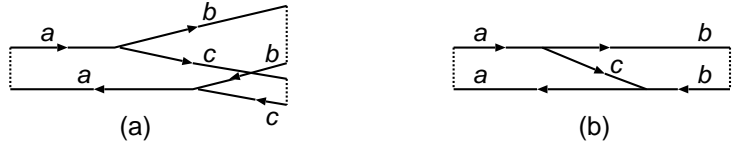,height=2cm}
\end{center}
\vspace{-.3cm}
\caption[.]{The diagram representation of the inclusive spectrum 
(5) (a), and (7) (b).}
\end{figure}
We will consider first the $\qb_{c}$-integrated spectrum. For 
the sake of generality we assume that 
all the particles are charged in this case. Later we will give the formula for the 
totally inclusive spectrum when at least one of the final particles has a zero charge.
For the $\qb_{c}$-integrated case the transverse density matrix for the 
final particle $c$ is given by a $\delta$-function, 
and taking advantage of the relation 
\beq
\int \!d\r_{2}K(\r_{2},z_{2}|\r_{1},z_{1})
K^{*}(\r_{2},z_{2}|\r_{1}^{'},z_{1})=\delta(\r_{1}-\r_{1}^{'})
\label{eq:6}
\eeq
one can transform
the graph of Fig.\,1a into that of Fig.\,1b.
The corresponding analytical
expression reads
\bea
\frac{d^{3}\sigma}{dx d\qb_{b}}=\frac{2}{(2\pi)^{2}}
\mbox{Re}\!\!
\int\!\!
d\r_{b,f}d\r_{b,f}^{'}d\r_{b}d\r_{b}^{'}
d\r_{a}d\r_{a}^{'}d\r_{a,i}d\r_{a,i}^{'}
\exp[-i\qb_{b}(\r_{b,f}-\r_{b,f}^{'})]
\!\int
\limits_{z_{i}}^{z_{f}}\!\!dz_{1}\!\!\int\limits_{z_{1}}^{z_{f}}\!\!dz_{2}
\nonumber \\
\times 
\,gS_{b}(\r_{b,f},\r_{b,f}^{'},z_{f}|
\r_{b},\r_{b}^{'},z_{2})
M(\r_{b},\r_{b}^{'},z_{2}|
\r_{a},\r_{a}^{'},z_{1})
S_{a}(\r_{a},\r_{a}^{'},z_{2}|
\r_{a,i},\r_{a,i}^{'},z_{i})\,,
\,\,\,\,\,\,\,\,\,\,\,\,\,\,\,\,
\label{eq:7}
\eea
where 
\beq
S_{i}(\r_{2},\r_{2}^{'},z_{2}|
\r_{1},\r_{1}^{'},z_{1})=
\langle 
K_{i}(\r_{2},z_{2}|\r_{1},z_{1})
K_{i}^{*}(\r_{2}^{'},z_{2}|\r_{1}^{'},z_{1})
\rangle
\label{eq:8}
\eeq
is the evolution operator for the transverse density matrix,
and the factor $M$ is given by
\beq
M(\r_{2},\r_{2}^{'},z_{2}|
\r_{1},\r_{1}^{'},z_{1})=
\langle 
K_{b}(\r_{2},z_{2}|\r_{1},z_{1})
K_{c}(\r_{2}^{'},z_{2}|\r_{1},z_{1})
K_{a}^{*}(\r_{2}^{'},z_{2}|\r_{1}^{'},z_{1})
\rangle\,.
\label{eq:9}
\eeq
We assume that the target density does not 
depend on $\r$. Then a considerable part of calculations can be 
done analytically.
In (\ref{eq:8}), (\ref{eq:9}) 
we represent the Green's functions in 
the path integral form. In the corresponding path integral formulas for 
$S_{i}$ and $M$ the interaction of the particles with the target potential 
after averaging over the target states turns out to be transformed into the 
interaction between trajectories described by the Glauber absorption
factors. 
For $S_{i}$ the corresponding absorption 
cross section is given by
the dipole cross section $\sigma_{i\bar{i}}$ 
of interaction with the medium constituent of $i\bar{i}$ 
pair. The absorption factor for $M$ involves 
the three-body cross section $\sigma_{\bar{a}bc}$
depending on the relative transverse vectors 
$\ta_{bc}=\r_{b}-\r_{c}$ and $\ta_{ab}=\r_{a}-\r_{b}$.
The factor $S_{i}$ can be evaluated analytically.
The corresponding formulas are given in \cite{BGZ3,BGZ1}. 
The factor $M$ after the analytical path integration over the center-of-mass coordinates
can be expressed through the Green's function $K_{bc}$ describing
the relative motion of the particles $b$ and $c$
in a fictitious $\bar{a}bc$ system. The formula for $M$ can be obtained 
from that given in \cite{BGZ1} by replacing the final transverse coordinate 
$\r_{2}$ by $\r_{2}^{'}$ for the particle $c$.
The expression for the probability of the $a\rightarrow bc$ transition
at a given impact parameter
which we obtain integrating  
analytically  over all the transverse coordinates
(except for $\ta_{b}=\r_{b,f}-\r_{b,f}^{'}$) 
in (\ref{eq:7}) has the form
\bea
\frac{d^{3}P}{dx d\qb_{b}}=\frac{2}{(2\pi)^{2}}
\mbox{Re}\!\!
\int\!\!
d\ta_{b}\,\exp(-i\qb_{b}\ta_{b})
\!\!\int
\limits_{z_{i}}^{z_{f}}\!\!dz_{1}\!\!\int\limits_{z_{1}}^{z_{f}}\!\!dz_{2}
\,g\Phi_{b}(\ta_{b},z_{2})
K_{bc}(\ta_{b},z_{2}|0,z_{1})
\Phi_{a}(\ta_{a},z_{1})
\,,
\label{eq:10}
\eea
where
\beq
\Phi_{a}(\ta_{a},z_{1})=
\exp\left[-\frac{\sigma_{a\bar{a}}(\ta_{a})}{2}
\int\limits_{z_{i}}^{z_{1}}\!dz n(z)\right]\,,
\,\,\,\,\,\,\,\,
\Phi_{b}(\ta_{b},z_{2})=
\exp\left[-\frac{\sigma_{b\bar{b}}(\ta_{b})}{2}
\int\limits_{z_{2}}^{z_{f}}\!dz n(z)\right]
\label{eq:11}
\eeq
are the eikonal initial- and final-state absorption
factors,\footnote{
Note that appearance of the eikonal absorption
factors in (\ref{eq:10}) is a nontrivial consequence 
of the specific form of the evolution operators
$S_{a,b}$ \cite{BGZ3}, and is not connected with applicability
of the eikonal approximation in itself.}
$\ta_{a}=x\ta_{b}$.
The Hamiltonian for the Green's function $K_{bc}$ reads
\beq
H_{bc}=-\frac{1}{2\mu_{bc}}
\left(\frac{\partial}{\partial\ta_{bc}}\right)^{2}
-\frac{in(z)\sigma_{\bar{a}bc}(\ta_{bc},\ta_{ab})}{2}
-\frac{i}{L_{f}}\,,
\label{eq:12}
\eeq
where $\mu_{bc}=E_{a}x(1-x)$, 
$\ta_{ab}=-[\ta_{a}+(1-x)\ta_{bc}]$,
$L_{f}=2E_{a}x(1-x)/[m_{b}^{2}(1-x)+m_{c}^{2}x-m_{a}^{2}x(1-x)]$
is the so-called formation length. 
In (\ref{eq:11}), (\ref{eq:12})
$n(z)$ is the number density of the target.
If the target occupies the region 
$0<z<L$ one can drop in 
(\ref{eq:10}) the contribution
from the configurations with $z_{1,2}<0$ and $z_{1,2}>L$. 
This follows from the relation
for the Green's function $K_{bc}^{0}$ for the Hamiltonian (\ref{eq:12}) in vacuum
\beq
\mbox{Re}
\int\limits_{0}^{\infty}\!
dz
K_{bc}^{0}(\ta,z|0,0)=
\mbox{Re}\int\limits_{0}^{\infty}\!dz
\left(\frac{\mu_{bc}}{2\pi iz}\right)
\exp\left\{i\left[\frac{\mu_{bc}\ta^{2}}{2z}-\frac{z}{L_{f}}\right]\right\}
=0\,.
\label{eq:13}
\eeq
This allows one to rewrite (\ref{eq:10}) 
in another form
\bea
\frac{d^{3}P}{dx d\qb_{b}}=\frac{2}{(2\pi)^{2}}
\mbox{Re}
\int\!
d\ta_{b}\,\exp(-i\qb_{b}\ta_{b})\,\,\,\,\,\,\,\,\,\,\,\,\,\,\,\,\,\,\,\,\,\,
\,\,\,\,
\nonumber\\
\times
\int
\limits_{z_{i}}^{z_{f}}\!dz_{1}\int\limits_{z_{1}}^{z_{f}}\!dz_{2}
\,g\left\{
\Phi_{b}(\ta_{b},z_{2})
\left[K_{bc}(\ta_{b},z_{2}|0,z_{1})-
K_{bc}^{0}(\ta_{b},z_{2}|0,z_{1})\right]
\Phi_{a}(\ta_{a},z_{1})
\right.
\nonumber\\
+\left.
\left[\Phi_{b}(\ta_{b},z_{2})-1\right]
K_{bc}^{0}(\ta_{b},z_{2}|0,z_{1})
\left[\Phi_{a}(\ta_{a},z_{1})-1\right]
\right\}
\label{eq:14}
\eea
which demonstrates explicitly that
the configurations with $z_{1,2}<0$ and $z_{1,2}>L$ do not contribute to
the radiation rate. Equations (\ref{eq:10}), (\ref{eq:14}) establish the
theoretical basis for evaluation of the transverse momentum dependence
of the LPM effect. 

The integration over $\qb_{b}$ in (\ref{eq:14})
gives the $x$-spectrum
\bea
\frac{dP}{dx}=2\mbox{Re}
\int
\limits_{z_{i}}^{z_{f}}\!dz_{1}\int\limits_{z_{1}}^{z_{f}}\!dz_{2}
\,g
\left[K_{bc}(0,z_{2}|0,z_{1})-
K_{bc}^{0}(0,z_{2}|0,z_{1})\right]
\label{eq:15}
\eea
which we obtained earlier in \cite{BGZ1}.
There it has been derived using
the unitarity connection between 
the probability of the $a\rightarrow bc$ transition 
and the radiative correction to the $a\rightarrow a$ transition. 
The latter is described by the diagram of Fig.\,2a which in turn
using (\ref{eq:6}) can be transformed into the graph of Fig.\,2b.
It can also be obtained directly from 
the graph of Fig.\,1b after integrating
over $\qb_{b}$.
\footnote{
The diagram of Fig.\,2b gives only the term $\propto K_{bc}$ in 
(\ref{eq:15}) (the corresponding integral is divergent in itself). 
Nonetheless, it yields the same result as (\ref{eq:15}). 
Indeed, by adding and subtracting the contribution from the configurations 
$z_{1}<z_{f}<z_{2}$, one can represent the contribution of the vacuum term
as a sum of the imaginary term connected with the radiative correction to $m_{a}$ 
(which $\propto (z_{f}-z_{i})$) and the real term related to
the wavefunction renormalization. The latter comes from the 
configurations $z_{1}<z_{f}<z_{2}$.
This boundary effect
is absent if the coupling vanishes at large $|z|$. 
In this case the vacuum term in (\ref{eq:15}) does not affect the $x$-spectrum.
It is, however, convenient
to keep the vacuum term to simplify the troublesome $z$-integration
in (\ref{eq:15}). Again, it allows one to use a constant coupling.}
\begin{figure}[h]
\begin{center}
\psfig{file=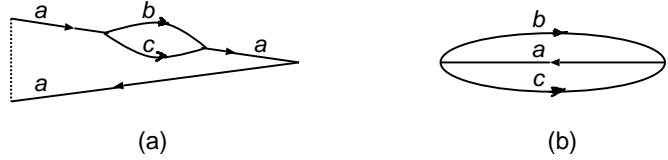,height=3cm}
\end{center}
\vspace{-.3cm}
\caption[.]{The diagram representation of the radiative correction 
to the probability of $a\rightarrow a$ transition.}
\end{figure}

In the low density limit (\ref{eq:14}) can be written
through the light-cone 
wavefunction $\Psi_{a}^{bc}$
\bea
\frac{d^{3}P}{dx d\qb_{b}}=
\frac{1}{(2\pi)^{2}}\int\!
d\ta d\ta^{'} \exp(-i\qb_{b}\ta^{'})
\Psi_{a}^{bc*}(x,\ta-\ta^{'})T\sigma_{\bar{a}bc}(\ta_{bc},\ta_{ab})
\Psi_{a}^{bc}(x,\ta)
\,,
\label{eq:16}
\eea
where
$\ta_{bc}=\ta$, $\ta_{ab}=-[(1-x)\ta +x\ta^{'}]$,
$T=\int dzn(z)$.
This formula can be obtained from (\ref{eq:14}) taking 
advantage of the representation for $\Psi_{a}^{bc}$
through $K_{bc}^{0}$ 
$$
\Psi_{a}^{bc}(x,\ta)=\frac{i\lambda}{4E_{a}\sqrt{\pi x(1-x)}}
\int\limits_{-\infty}^{0}
dz K_{bc}^{0}(\ta,0|0,z)
$$
established in \cite{BGZ_SLAC}.
Being divided by $T$ (\ref{eq:16}) gives
a convenient formula for the Bethe-Heitler cross section 
in terms of the light-cone wavefunction. It worth noting
that (\ref{eq:16}) (and (\ref{eq:10}), (\ref{eq:14}) as well) is valid
if one can neglect the transverse motion effects on the scale
of the medium constituent size. 
This assumes that the typical value of $|z_{2}-z_{1}|$ in (\ref{eq:10}), 
(\ref{eq:14}), which can be regarded as the formation length 
associated with the $a\rightarrow bc$
transition at a given $\qb_{b}$, $L_{f}(q_{b})$, is  
much larger than the size of the medium constituent. 
If the LPM effect
is not very strong the $L_{f}(q_{b})$ can be estimated replacing 
$m_{b,c}^{2}$ by $m_{b,c}^{2}+\qb_{b}^{2}$ in the above formula for $L_{f}$.
Note that for $L_{f}(q_{b})\gg L$
the radiation rate can be written
through $\Psi_{a}^{bc}$ for arbitrary target
density. In this case the target acts as a single
scattering center
and (\ref{eq:14}) can be written  in a form like (\ref{eq:16})
but with the product $T\sigma_{\bar{a}bc}$ being replaced by 
$
2\left\{1-\exp\left[-\frac{1}{2}T\sigma_{\bar{a}bc}\right]
\right\}\,.
$
This representation 
generalizes the formula
for the $p_{T}$-integrated spectrum derived in \cite{NPZ}.

In general case one can estimate the radiation rate
using the parametrizations
$\sigma_{i\bar{i}}=C_{ii}\ta_{i}^{2}$,
$
\sigma_{\bar{a}bc}=C_{ab}\ta_{ab}^{2}+
C_{bc}\ta_{bc}^{2}+C_{ca}\ta_{ca}^{2}
%\label{eq:15}
%\eeq
$
(here $\ta_{ca}=-(\ta_{ab}+\ta_{bc})$).
Then the Hamiltonian (\ref{eq:12}) takes the oscillator
form with the frequency
$
\Omega(z)=\frac{(1-i)}{\sqrt{2}}
\left[\frac{n(z)C(x)}{E_{a}x(1-x)}\right]^{1/2}\,,
%\label{eq:16}
$
where 
$C(x)=C_{ab}(1-x)^{2}+
C_{bc}+C_{ca}x^{2}$.
The Green's function for the 
oscillator Hamiltonian 
with the $z$-dependent frequency
can be written in the form
\beq
K_{osc}(\ta_{2},z_{2}|\ta_{1},z_{1})=
\frac{\gamma(z_{1},z_{2})}{2\pi i}
\exp\left\{i\left[\alpha(z_{1},z_{2})\ta_{2}^{2}
+\beta(z_{1},z_{2})\ta_{1}^{2}-\gamma(z_{1},z_{2})\ta_{1}\ta_{2}
\right]\right\}
\,.
\label{eq:17}
\eeq
The functions $\alpha$, $\beta$ and $\gamma$ in (\ref{eq:17}) can be
evaluated in the approach of Ref. \cite{Jpsi}.
Then we can integrate analytically over $\ta_{b}$
in (\ref{eq:10}), and represent the radiation rate as 
\bea
\frac{d^{3}P}{dx d\qb_{b}}=\frac{1}{(2\pi)^{2}}\mbox{Re}
\int\limits_{z_{1}<z_{2}}
\!dz_{1} dz_{2}\,g\,
\frac{\gamma(z_{1},z_{2})}{Q(z_{1},z_{2})}
\exp\left[-\frac{i\qb^{2}_{b}}{4Q(z_{1},z_{2})}
+\frac{i(z_{1}-z_{2})}{L_{f}}\right]
\,,
\label{eq:18}
\eea
where the factor $Q(z_{1},z_{2})$ can be expressed
through the parameters $C_{ij}$, functions $\alpha$,
$\beta$, $\gamma$, and $n$. The formula 
for this factor is cumbersome to be presented here.

Consider now the case when the particle $a$ is produced in a medium
or at finite distance from a target. Equation (\ref{eq:10}) 
holds in this case as well but 
now $z_{i}$ equals the coordinate of the production point.
Given the representation (\ref{eq:10}) taking advantage of (\ref{eq:13}) one 
can obtain a formula similar to (\ref{eq:14}) but with 
$[\Phi_{a}-1]$ being replaced by $\Phi_{a}$ in the second term.
Note, however, that, due to infinite time required for the formation
of $\Psi_{a}^{bc}$,  equation (\ref{eq:16}) 
(and its analogue for arbitrary density at $L_{f}(q_{b})\gg L$) 
does not hold in this case.

Let us discuss briefly the totally inclusive radiation rate. It can be
evaluated almost in the same way as the $\qb_{c}$-integrated spectrum
if one of the final particles has a zero charge, as this occurs for 
the $e\rightarrow \gamma e$ transition in QED.
Consider the case when $e_{c}=0$. Since the particle $c$ does not 
interact with the medium the graph of Fig.\,1a can be transformed into
a graph like that of Fig.\,1b but with the propagator $K_{c}$ 
being connected to the lower $abc$ vertex through the density matrix of the particle $c$.
The corresponding formula (which is the analogue of (\ref{eq:10})) reads
\bea
\frac{d^{5}P}{dx d\qb_{b}d\qb_{c}}=\frac{2}{(2\pi)^{2}}
\mbox{Re}
\int\!
d\ta_{b}d\ta_{c}\,\exp\left[-i(\qb_{b}\ta_{b}+\qb_{c}\ta_{c})\right]
\nonumber\\
\times
\int
\limits_{z_{i}}^{z_{f}}\!dz_{1}\int\limits_{z_{1}}^{z_{f}}\!dz_{2}
\,g\Phi_{b}(\ta_{b},z_{2})
K_{bc}(\ta_{b}-\ta_{c},z_{2}|0,z_{1})
\Phi_{a}(\ta_{a},z_{1})
\,,
\label{eq:19}
\eea
where $\ta_{a}=x\ta_{b}+(1-x)\ta_{c}$. The $z$-integration in
(\ref{eq:19}) can also be written as in (\ref{eq:14}). In the 
low density limit and at $L_{f}(q_{b})\gg L$ the initial- and final-state 
interaction vanish. For this reason the analogue of (\ref{eq:16}) and a similar
equation for arbitrary density at $L_{f}(q_{b})\gg L$ which can be obtained
from (\ref{eq:19}) are valid even when all the particles are charged.

The generalization of the above results to the realistic QED and QCD
Lagrangians reduces to trivial replacements of the two- and 
three-body cross sections, and vertex factor $g$. The latter, 
due to spin effects
in the vertex $a\rightarrow bc$, 
becomes an operator. The corresponding formulas 
are given in \cite{BGZ1,YAF}.

The formalism  developed can be applied to many 
problems. In particular, in QCD this approach can be used for evaluation of
high-$p_{T}$ hadron spectra, the $p_{T}$-dependence
of Drell-Yan pairs and heavy quarks production in $hA$-collisions, 
angular dependence of the parton energy
loss in hot QCD matter produced in $AA$-collisions. It is also
of interest for study the initial condition 
for quark-gluon plasma in $AA$-collisions.
Some of these problems will be discussed in further publications.\\

%\section*{Acknowledgments}
I would like to thank N.N. Nikolaev and D. Schiff for
discussions. I am grateful to J.~Speth for the hospitality
at FZJ, J\"ulich, where this work was completed.   
This work was partially supported by the grants INTAS
96-0597 and DFG 436RUS17/11/99.

%\pagebreak

\end{document}